# Universal relationship between crystallinity and irreversibility field of MgB$_2$

A. Yamamoto, J. Shimoyama, S. Ueda, Y. Katsura, I. Iwayama, S. Horii and K. Kishio

*Department of Superconductivity, University of Tokyo, 7-3-1 Hongo, Bunkyo-ku, Tokyo, 113-8656, Japan*

(April 30, 2005)

The Relationship between irreversibility field, $H_{irr}$, and crystallinity of MgB$_2$ bulks including carbon substituted samples was studied. The $H_{irr}$ was found to increase with an increase of FWHM of MgB$_2$ (110) peak, which corresponds to distortion of honeycomb boron sheet, and their universal correlation was discovered even including carbon substituted samples. Excellent $J_c$ characteristics under high magnetic fields were observed in samples with large FWHM of (110) due to the enhanced intraband scattering and strengthened grain boundary flux pinning. The relationship between crystallinity and $H_{irr}$ can explain the large variation of $H_{irr}$ for MgB$_2$ bulks, tapes, single crystals and thin films.

The superconductivity at 39 K discovered in MgB$_2$ has attracted great interest in its practical applications.[1] Now, MgB$_2$ is showing higher $H_{c2}$ than that of conventional NbTi and Nb$_3$Sn and becoming the first promising metallic superconductor applicable at ~20 K. In particular, very high upper critical field, $H_{c2}(0)$, exceeding ~70 T has been reported for moderately alloyed thin films,[2-4] while the highest $H_{c2}(0)$ still remains ~30 T for wires, tapes, and bulks.[5-8] Among the numerous approaches to improve flux pinning properties of MgB$_2$ by chemical method, doping of carbon[4,6,8,9] or carbon contained compounds, such as SiC[5,7] and B$_4$C,[10] is well known to be most effective, while detailed mechanisms of the improved $J_c$ characteristics have not been well clarified yet. On the other hand, it is also well known that the superconducting properties of MgB$_2$, such as $T_c$, $H_{c2}$, $J_c$, and normal state resistivity are largely influenced by the purity and size of starting powders of Mg and B[11] as well as the heating conditions.[12]

Compared to the other metallic superconductors, the irreversibility field, $H_{irr}$, of MgB$_2$ is apparently lower than $H_{c2}$ due to weak flux pinning. A typical relationship, $H_{irr}(T) \sim 0.5 H_{c2}(T)$, has been observed for the undoped MgB$_2$ bulks.[12-13] Furthermore, the predominant pinning mechanism in MgB$_2$ is still controversial, however, grain boundaries are believed to act as pinning centers.[14-17] Grain size and impurity parameter $\alpha$, *i.e.*, moderate dirtiness of the samples, are crucial factors of grain boundary pinning.[18-19]

In our previous study, crystallinity dependent $H_{c2}$ and $H_{irr}$ were found for MgB$_2$ bulks synthesized by various heating conditions, which have various full-width at half-maximum (FWHM) values derived from powder X-ray diffraction (XRD) peaks.[12] Among various XRD peaks, the FWHM of the (110) reflection corresponding to the in-plane disorder was found to be strongly dependent on the samples. The degradation of crystallinity originates from disordered crystal lattice caused by various types of lattice defects or intragranular precipitates.[20] These disorders in grains certainly enhance the interband and intraband scattering in σ, π bands, and moderate increase of intraband scattering directly results in an improvement of $H_{c2}$.[4-5,21]

In this letter, we demonstrate the universal and strong relationship between crystallinity and $H_{irr}$ of MgB$_2$ including carbon doped samples, which may explain the wide variation of $H_{irr}$ among the wires, tapes, bulks, thin films and single crystals. Our results conclusively suggest that the disorder in honeycomb boron sheet is strongly related to the flux pinning strength through the change in $H_{c2}$.

The polycrystalline MgB$_2$ bulks were synthesized by the previously developed PICT method.[22] Powder mixture of Mg and B sealed in SUS316 tube was heated in an evacuated quartz ampoule. MgB$_2$ samples with different crystallinity were obtained by control of heating conditions (500-1100°C, 1-1200 h). Carbon substituted MgB$_2$ samples were fabricated by addition of B$_4$C or SiC. Crystallinity of the samples was evaluated by powder X-ray diffraction analysis using the Cu-$K_\alpha$ radiation.

Figure 1 shows powder XRD patterns of three typical MgB$_2$ bulks, #1 (undoped, heated at 900°C for 60 h), #2 (undoped, heated at 650°C for 3 h) and #3 (SiC 20mol% doped, heated at 600°C at 24 h). The FWHM of both (002) and (110) peaks for #2 are larger than those of #1.

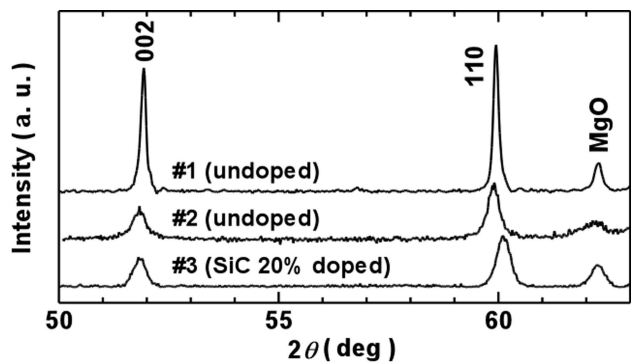

FIG. 1. Powder X-ray diffraction patterns of MgB$_2$ bulks, #1 (undoped, heated at 900°C for 60 h), #2 (undoped, heated at 650°C for 3 h) and #3 (SiC 20% doped, heated at 600°C for 24 h).

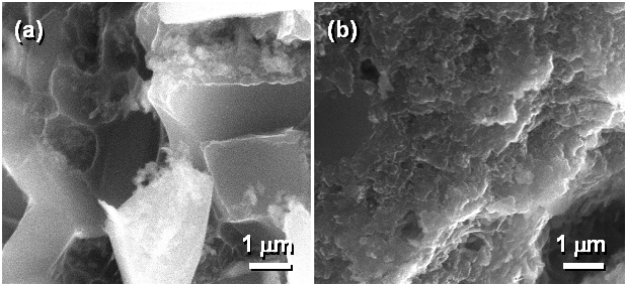

FIG. 2. Secondary electron images of the fractured surface of the MgB$_2$ bulk; (a) #1 (undoped, heated at 900°C for 60 h); (b) #3 (SiC 20% doped, heated at 600°C for 24 h).

In #3, the (110) peak is located higher angle than those of #1 and #2, meaning the shrinkage of $a$-axis length by carbon substitution for boron site. The FWHM of (110) for the #3 was largest among these three samples, while difference in FWHM of (002) peaks between #2 and #3 was small. This result indicates that carbon substitution introduces mainly in-plane lattice disorder in the boron sheet.

SEM observation revealed that samples heated above 900°C for long time have large MgB$_2$ grains with several μm in size as shown in Fig. 2(a), while mean grain size of other samples was ~300nm independent of heating conditions and doping levels of SiC or B$_4$C as shown in Fig. 2(b).

Magnetic field dependences of $J_c$ estimated from magnetization hysteresis loops based on the Bean Model for samples with different crystallinity are shown in Figure 3. In undoped samples, $J_c$ under high magnetic fields was systematically improved with an increase of FWHM of (110) peak, while $J_c$ under low fields did not largely change except two samples having large grains. Further improved $J_c$ under high fields was observed in the SiC doped samples, and very high $J_c$ of 2,500 A/cm$^2$ at 20 K under 5 T, which is comparable to the recently reported highest $J_c$ by Dou et al.,[7] was obtained for #3. Poor $J_c$ properties observed in samples that were heated at high-temperatures and exhibit sharp (110) peaks are believed to be due to largely grown grains with high crystallinity, which reduce pinning strength at the grain boundaries. Therefore, the flux pinning properties of #1 is rather similar to those of single crystals in the clean-limit.[23]

The $H_{irr}$ defined by a $J_c$ criterion of 10$^3$ A/cm$^2$ were plotted as a function of the FWHM of (110) peaks in Figure 4. It is clear that there is a strong correlation between FWHM of (110) peak and $H_{irr}$. On the other hand, less prominent correlation between $H_{irr}$ and FWHM of (002) peak was confirmed. These mean that introduction of disorders to the $ab$-plane, i.e., distortion in honeycomb boron sheet, is essentially effective to enhance $H_{irr}$.

In a previous study, we have reported that degradation of crystallinity is also correlated with the enhancement of $H_{c2}$.[12] The improved $H_{c2}$ is a reasonable result of a reduced coherence length $\xi$ due to the enhanced intraband scattering. Degradation of crystallinity is originated from the lattice distortion and intragranular precipitates. These are believed to increase the intraband scattering, shorten the electron mean free path $l$ and $\xi$ according to the equation $1/\xi = 1/l + 1/\xi_0$. Actually, higher resistivity of 110 μΩcm at 40 K was observed in the dirty #2, while 3.8 μΩcm for the clean #1. The enhancement of $H_{irr}$ with an increase in FWHM of (110) peak is explained as a result of strengthened grain boundary pinning. Yetter et al. pointed out that electron scattering pinning at grain boundary (so called $\delta\kappa$ pinning) is strongly dependent upon the purity of the sample.[18] The pinning force at grain boundaries ($f_{GB}$) is increasing with an increase of impurity parameter $\alpha = 0.882\xi_0/l$ in low $\alpha$ region. Since $f_{GB}$ is proportional to the inverse of grain size and lattice strain shortens $l$ and increases $\alpha$, both small crystal size and lattice distortion resulting in broad XRD peaks are considered to enhance the grain boundary pinning.

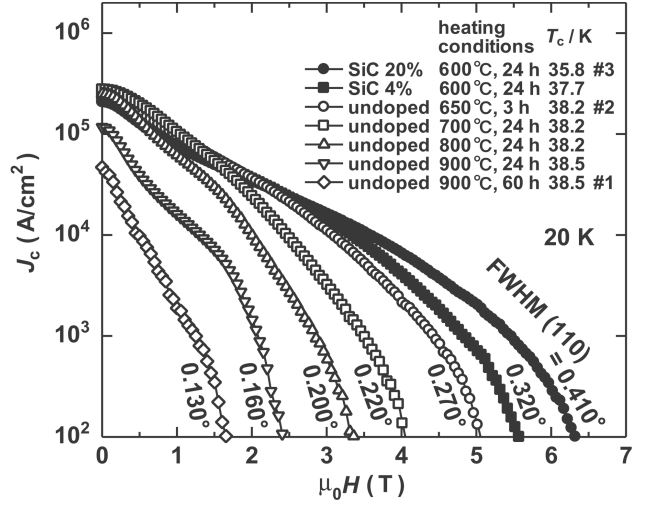

FIG. 3. Magnetic field dependence of $J_c$ at 20 K for undoped and SiC doped MgB$_2$ bulks with various crystallinity.

Higher $H_{irr}$ with larger FWHM observed in B$_4$C, SiC doped samples with the same trend for undoped samples directly suggests that improvement of $H_{irr}$ by carbon substitution is attributed to the enhanced grain boundary pinning. Since substituted carbon atoms do not act as strong pinning centers,[24] $H_{irr}$ is suggested to be determined by the same pinning mechanism with undoped MgB$_2$. Increased intraband scattering in $\sigma$, $\pi$ bands introduced by carbon substitution, which enhances $H_{c2}$, also improves the grain boundary pinning. The carbon substitution effect on the enhanced $J_c$ properties might be simply explained by the introduction of lattice disorders into MgB$_2$ grains.

The relationship between crystallinity and $H_{irr}$ can explain the largely varied $H_{irr}$ of MgB$_2$ samples including tapes, single crystals and thin films. For example, single crystals with sharp XRD peaks and no grain boundary showed low $H_{irr}$,[23] and the dirty film with large FWHM value over 1°, which has nano-MgB$_2$ grains containing large amount of MgO nano-precipitates, showed high

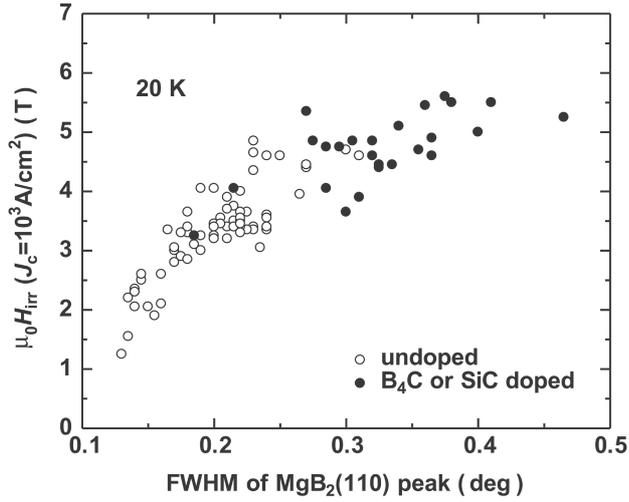

FIG. 4. Relationship between irreversibility field at 20 K and FWHM of MgB$_2$ (110) peak for undoped and B$_4$C or SiC doped MgB$_2$ bulks. Samples with $T_c$ >35 K were selected for B$_4$C or SiC doped bulks.

$H_{irr}$.[2]

In summary, a universal relationship between $H_{irr}$ and crystallinity was found for MgB$_2$ bulks even including carbon substituted samples. Strong $H_{irr}$ dependence of FWHM of MgB$_2$ (110) in-plane peak in both undoped and carbon substituted samples implies that distortion of honeycomb boron sheet is directly associated with the intraband scattering resulting in an enhancement of grain boundary flux pinning. Microstructural analyses for samples with various crystallinity will provide more essential information for designing MgB$_2$ materials with high $H_{c2}$ and strong flux pinning.

The authors are grateful to Drs. T. Machi and N. Chikumoto at ISTEC-SRL for high field magnetization measurements. This study was partially supported by Special Coordination Funds of the Ministry of Education, Culture, Sports, Science and Technology of the Japanese government.